\documentclass[11pt]{amsart}
\usepackage{amsmath,amsthm,amsfonts,amssymb,times,bm}
\usepackage{latexsym}
\usepackage{mathtools}

\usepackage{mathrsfs} 

\numberwithin{equation}{section}  

\DeclareMathAlphabet{\curly}{U}{rsfs}{m}{n}  

\theoremstyle{remark}
\newtheorem{remark}{Remark}
\newtheorem{eexample}{Example}
\theoremstyle{plain}
\newtheorem{proposition}{Proposition}

\newtheorem{theorem}{Theorem}

\newtheorem{corollary}{Corollary}
\newtheorem{definition}{Definition}

\newcommand{\D}{\mathsf D}
\newcommand{\R}{\mathsf R}
\newcommand{\Adj}{\text{Adj}}
\newcommand{\Prob}{\mathbb P}

\makeatletter
\renewcommand{\pmod}[1]{\allowbreak\mkern7mu({\operator@font mod}\,\,#1)}
\makeatother

\begin{document}

\title{Privacy-Preserving Nonlinear Observer Design Using Contraction Analysis}

\author{Jerome Le Ny}
\address{Department of Electrical Engineering \& GERAD\\ Polytechnique Montreal\\ C.P. 6079, succursale Centre-ville\\ Montreal, QC H3C-3A7\\ Canada}
\email{jerome.le-ny@polymtl.ca}

\thanks{This work was supported by NSERC under Grant RGPIN-435905-13. 
A preliminary version of this paper was presented at CDC 2015 \cite{LeNy:CDC15:contractionDP:1}.}



\begin{abstract} Real-time information processing applications such as those enabling a more intelligent infrastructure  
are increasingly focused on analyzing privacy-sensitive data obtained from individuals.
To produce accurate statistics about the habits of a population of users of a system, this data might need to 
be processed through model-based estimators. Moreover, models of population dynamics, originating for example 
from epidemiology or the social sciences, are often necessarily nonlinear.   
Motivated by these trends, this paper presents an approach to design nonlinear privacy-preserving model-based observers, 
relying on additive input or output noise to give differential privacy guarantees to the individuals providing the input data.
For the case of output perturbation, contraction analysis allows us to design convergent observers as well as set the level 
of privacy-preserving noise appropriately. Two examples illustrate the approach: estimating the edge formation probabilities in a   
dynamic social network, and syndromic surveillance relying on an epidemiological model.
\end{abstract}

\maketitle



\section{Introduction}

The possibility to analyze vast amounts of personal data capturing information about the activities of private 
individuals is a foundational principle behind many current technology-driven trends such as the ``Internet of Things'', 
electronic biosurveillance systems, or developing an intelligent infrastructure enabling smart cities. In many respects however, 
the data collection practices envisioned to operate these systems often go against basic privacy 
rights \cite{Warren:1890HLR:PrivacyRight:1}.
Concerns about the acquisition and use of personal data by companies and governments, 
e.g., for potential price and service discrimination, are rising \cite{McDaniel:Sec09:smartGridSP, 
PCAST14:privacy:1, Markey:report15:connectedCars:1}, 
and could lead to people rejecting these technologies despite their suggested benefits.
Rigorous privacy-preserving data analysis methodologies are needed to support regulations and allow
people to appropriately trade off the privacy risks they increasingly incur with the benefits they can expect
in return.

Typically, large-scale monitoring and control systems only require aggregate statistics computed
from personal data streams, e.g., a dynamic map showing road traffic conditions built from location traces
sent by smartphones, or an estimate of power consumption in a neighborhood updated using smart meter data
from individual homes.
Aggregation is beneficial to privacy, but past examples have shown that it is not sufficient 
to a priori rule out the possibility of significant privacy breaches 
\cite{Narayanan08_netflixBreach:1, Calandrino11_privacyAttackCollabFilt:1, Wilson05_tracking:1}. 
Privacy attacks are often \emph{linkage attacks}, where some newly published information is 
combined with other available data to make new inferences about specific individuals, and predicting at system design time
how such attacks could be carried out is difficult.
Yet, as explained below, it is still possible to compute aggregate statistics with formal privacy guarantees for the individuals 
from whom the data originates, which could help alleviate some of the justified concerns and encourage 
wider adoption of certain pervasive sensing and control systems. 

Various information theoretic definitions have been proposed to capture quantitatively the concept of privacy 
and are potentially applicable to the processing of data streams in real-time \cite{2013_SPprivacySpecialissue:1}. 
In this paper, we focus on the notion of \emph{differential privacy}, which originates from the database and cryptography 
literature \cite{Dwork06_DPcalibration:1}. Intuitively, a differentially private mechanism publishes information 
about a dataset in a way that is not too sensitive to a single individual's data. 
As a result, an individual receives a guarantee that by providing her data, she will
not drastically change the ability of a third party to make new inferences about her.
Previous work has considered the design of linear filters processing sensitive time series data with 
differential privacy guarantees \cite{Dwork10_DPcounter:1, Chan11_DPcontinuous:1, LeNy_Allerton2012_DPKF:1, Bolot13_privateDecayedSums:1,  
LeNyDP2012_journalVersion:1, LeNy:TAC18:MIMOdpef, McGlinchey:IFAC17:positive}.  
The problem studied in this paper is that of designing privacy-preserving nonlinear model-based estimators, 
which to the best of our knowledge has not been studied in a general setting before.
A convenient way of achieving differential privacy for
an estimator is to bound its so-called \emph{sensitivity} \cite{Dwork06_DPcalibration:1}, a form of incremental system gain 
between the private input signals and the published output \cite{LeNyDP2012_journalVersion:1}. 
Various tools could be used for this purpose, and here we rely on contraction 
analysis \cite{Lohmiller98_contraction:1, Sontag10_contraction:1, Forni14_contraction:1, Angeli00_contraction:1}.

The rest of the paper is divided as follows. Section \ref{section: statement} presents the problem statement formally,
provides a brief introduction to the notion of differential privacy, and describes and compares privacy-preserving data analysis mechanisms 
with input and output perturbations.
In Section \ref{section: contraction} we discuss some fundamental results in contraction analysis and 
present a type of ``Input-to-State Stability''   
property of contracting systems similar to the one proved in \cite{Sontag10_contraction:1} but stated 
here for discrete-time systems. This property is used in Section \ref{section: DP observers} to design differentially
private observers with output perturbation. 
The methodology is illustrated via two examples involving the analysis of dynamic data originating from private individuals.
In Section \ref{section: DSBM}, we consider the problem
of estimating link formation probabilities in a dynamic social network, with a nonlinear measurement model. 
In Section \ref{ref: syndromic surveillance}, we consider a nonlinear epidemiological model and design a differentially
private estimator of the proportion of susceptible and infectious people in a population, assuming a syndromic data source.

\emph{Notation:} 
In this paper, $\mathbb N := \{0,1,\ldots\}$ denotes the set of non-negative integers.
For $H:\mathsf X \to \mathsf Y$ a linear map between finite dimensional vector spaces $\mathsf X$ and $\mathsf Y$ 
equipped with the norms  $|\cdot|_\mathsf X$ and  $|\cdot|_\mathsf Y$ respectively, we denote by $\|H\|_{\mathsf X}^{\mathsf Y}$ 
its induced norm, so that $|H x|_\mathsf Y \leq \|H\|_{\mathsf X}^{\mathsf Y} \, |x|_\mathsf X$, for all $x$ in $\mathsf X$.
If $\mathsf X = \mathsf Y$ and both spaces are equipped with the same norm $|\cdot|_\mathsf X$, 
we simply write $\|\cdot\|_\mathsf X$. 
For $1 \leq p < \infty$, the $p$-norm on $\mathbb R^n$, denoted $| \cdot |_p$, is defined as 
$| v |_p := \left(\sum_{i=1}^n |v_i|^p\right)^{1/p}$, and $| v |_\infty := \max_{1 \leq i \leq n} |v_i|$.  
For $v=\{v_k\}_{k \in \mathbb N}$ a vector-valued discrete-time signal, where $v_k \in \mathbb R^n$ has components $\{v_{k,i}\}_{i=1}^n$,
the $\ell_p$ signal norm is $\| v \|_p = \left( \sum_{k=0}^\infty \sum_{i=1}^n | v_{k,i} |^p \right)^{1/p} = \left( \sum_{k=0}^\infty |v|_p^p \right)^{1/p}$
for $1 \leq p < \infty$, and $\| v \|_\infty = \sup_{k \geq 0} |v_k|_\infty$.
We use $\text{diag}(v)$ to denote a diagonal matrix with the components of the vector $v$ on the diagonal.
For $P$ a symmetric matrix, $P$ positive definite is denoted $P \succ 0$ and $P$ positive semidefinite is denoted $P \succeq 0$.
For $P \succeq 0$, we denote its (unique) positive semi-definite square root as $P^{1/2}$, i.e., $P = P^{1/2}P^{1/2}$.
For $P$, $Q$ symmetric matrices, $P \succeq Q$ means $P - Q \succeq 0$, and $P \preceq 0$ means $-P \succeq 0$.
The expressions ``if and only if'' and ``independent and identically distributed'' are abbreviated as iff and iid respectively.
$\mathcal C^1$ denotes the set of continuously differentiable functions. A class $\mathcal K$ function $\beta: \mathbb R_+ \to \mathbb R_+$
is a strictly increasing continuous function such that $\beta(0) = 0$.


\section{Problem Statement}	\label{section: statement}

\subsection{Observer Design}

Consider the problem of estimating a discrete-time signal denoted $x := \{x_k\}_{k \in \mathbb N}$, with
$x_k \in \mathsf X = \mathbb R^n$ for some integer $n$, which represents an aggregate state 
for a population of privacy-sensitive individuals.
For example, $x_k$ could be the density at period $k$ of drivers or pedestrians
at a finite number of spatial locations, the proportion of individuals infected by a disease in a population, etc.
We assume that $x_k$ cannot be perfectly observed, but that we can measure instead
a privacy-sensitive discrete-time signal $\{y_k\}_{k \in \mathbb N}$, with $y_k \in \mathsf Y = \mathbb R^m$ for some integer $m$, 
for which we have a state-space model of the form
\begin{align}	
x_{k+1} &= f_k(x_k) + w_k,  	\label{eq: model - 1} \\
y_k &= g_k(x_k) + v_k,		\label{eq: model - 2}
\end{align}
where $w_k, v_k$ are noise signals capturing modeling errors,
and $f_k$ and $g_k$ are $\mathcal C^1$ functions.
Our aim is to publish an estimate $z_k$ of $x_k$, 
computed from $y_k$ by an observer of the following form \cite{Sontag:SCL99:OSS} 
\begin{align}	\label{eq: observer form}
z_{k+1} = f_k(z_k) + h_k(z_k, y_k-g_k(z_k)),
\end{align}
with, for each $k$ in $\mathbb N$, $h_k: \mathsf X \times \mathsf Y \to \mathsf X$ a $\mathcal C^1$ function such that $h_k(x,0) = 0$.
We initialize \eqref{eq: observer form} with some estimate $z_0$ of $x_0$.  
Note that \eqref{eq: observer form} could describe an observer for a model  
\eqref{eq: model - 1}-\eqref{eq: model - 2} that has already been transformed under a suitable change of 
coordinates to a form that facilitates observer design, e.g., an observability canonical form 
\cite{Gauthier:book01:observability, Isidori:book17:controlLectures}.  
With straightforward modifications to our arguments, the ``prediction'' form \eqref{eq: observer form} could also be replaced by 
an observer using the most recent observations
\begin{align}	
z_0 &= \bar z_0 + h_0(\bar z_0, y_{0}-g_k(\bar z_0)), \quad \text{for some estimate $\bar z_0$ of $x_0$},\nonumber \\
z_{k+1} &= f_k(z_k) + h_{k+1}(z_k, y_{k+1}-g_k(f_k(z_k))), \; \text{for } k \geq 0.  \label{eq: observer form 2}
\end{align}

In the applications discussed later in the paper, the signal $y_k$ is collected from privacy-sensitive individuals, 
hence needs to be protected, in a sense defined below. 
On the other hand, the model \eqref{eq: model - 1}-\eqref{eq: observer form}, i.e.,
the functions $f_k$, $g_k$ and $h_k$, 
is assumed to be publicly available, or at least potentially known 
to any adversary trying to make inferences about $y$ based on $z$. 
The data aggregator wishes to publicly release the signal $z$ produced by (\ref{eq: observer form}). 
However, since $z$ depends on the sensitive signal $y$, we only allow the release of an approximate version 
of $z$ carrying certain privacy guarantees, which are presented formally in the next subsection.
As a result, it will emerge that the functions $h_k$ 
need to be carefully chosen to balance accuracy 
or convergence speed of the observer with the level of privacy offered.

\begin{remark}
We do not provide here nor use any model of the noise signals $w$ and $v$ in \eqref{eq: model - 1}, \eqref{eq: model - 2},   
which are simply introduced as a device to explain the discrepancy between any measured signal $y$ 
and the signals that can be predicted by a deterministic model $x_{k+1} = f_k(x_k), y_k = g_k(x_k)$.
\end{remark}

\begin{remark}
More generally, we might just want to publish an output $\chi_k(x_k)$,  
function of the state $x_k$. As explained below, this can be done  
by first obtaining a privacy-preserving estimate $\hat x$ of the signal $x$ and then publishing $\chi_k(\hat x_k)$, 
relying on the fact that sound privacy guarantees such as differential privacy are preserved by the final transformation through $\chi_k$.  
\end{remark}

\subsection{Differential Privacy}	\label{section: dp definition}

The published signal should provide an accurate estimate of $x$ under an additional constraint
that is not satisfied a priori by $z$ from \eqref{eq: observer form}, aiming at preserving the privacy 
of the individuals from which the measured signal $y$ originates. 
More precisely, we impose that the published signal be differentially private \cite{Dwork06_DPcalibration:1},
which requires adding artificial noise somewhere in the signal processing system to randomize the published output.  
A differentially private version of the observer (\ref{eq: observer form}) should produce a randomized output 
signal whose distribution is not too sensitive to certain variations associated with the effect of any individual's data 
on the signal $y$, input of the observer. 
The formal definition of differential privacy is given in 
Definition \ref{def: differential privacy original} below and requires that we specify the 
type of variations in $y$ that should be 
hard to detect from the published output.
This is done by defining a symmetric binary relation, called adjacency and denoted $\Adj$, 
on the space of datasets $\D$ of interest, here the space of signals $y$, so that two adjacent 
input signals $y$ and $\tilde y$ should produce (randomized) output signals with similar distributions.
It is possible to define different adjacency relations  \cite{LeNyDP2012_journalVersion:1}
to model different data analysis scenarios. In this paper, $y$ is assumed to represent a 
(possibly multi-dimensional) signal that already aggregates the data obtained from multiple users,
e.g., $y_k$ at a particular time period $k$ could be the number of people waiting in a hospital emergency room,
the total power consumption of a group of homes during that period, etc.
We then consider in particular the following adjacency relations between signals
\begin{align} \label{eq: adjacency relation 0}
&\Adj(y,\tilde y) \text{ iff } \|y - \tilde y\|_p \leq B_p, 
\end{align}
for $p = 1$ or $p = 2$ and some given fixed constant $B_p > 0$, as well as the more restrictive adjacency relation
\begin{align}		\label{eq: adjacency relation}  
&\Adj(y,\tilde y) \text{ iff } 
\exists k_0 \geq 0 \text{ s.t. } 
\begin{cases}
y_k = \tilde y_k, & k < k_0 \\
|y_k - \tilde y_k|_{p} \leq K \alpha^{k-k_0}, & k \geq k_0, 	
\end{cases}
\end{align}
where again $p = 1$ or $p = 2$ and $K>0$, $0 \leq \alpha<1$ are given fixed constants.  
In other words, we aim at 
hiding deviations in the signal $y$ 
(e.g., due to the contribution of one individual to the signal)
that are bounded in $p$-norm (relation \eqref{eq: adjacency relation 0}), or
more explicitly that can start at any time $k_0$ but then subsequently decrease geometrically (relation  \eqref{eq: adjacency relation}).
Note that even the more restrictive condition (\ref{eq: adjacency relation}) is much more general than the adjacency relation considered in 
some previous work on the design of a differentially private 
counter \cite{Dwork10_DPcounter:1, Chan11_DPcontinuous:1, Bolot13_privateDecayedSums:1}, 
where adjacent (scalar) signals can vary at a single time period by at most one.
In comparison, the adjacency condition (\ref{eq: adjacency relation}) greatly enlarges the set 
of signal deviations that can result from the presence of any individual and for which we 
provide guarantees (deviation at a single period is obtained for $\alpha = 0$). 
We can now state the definition of a differentially private mechanism, i.e., of a randomized map
from input to output signals.
\begin{definition}	\label{def: differential privacy original}
Let $\D$ be a space equipped with a symmetric binary relation denoted $\Adj$, and let $(\R, \mathcal M)$ 
be a measurable space, where $\mathcal M$ is a given $\sigma$-algebra over $\R$. 
Let $\epsilon, \delta \geq 0$. A randomized mechanism $M$ from $\D$ to $\R$ is
$(\epsilon, \delta)$-differentially private (for $\Adj$) if for all $d,d' \in \D$ such that $\Adj(d,d')$, 
we have
\begin{align}	\label{eq: standard def approximate DP original}
\Prob(M(d) \in S) \leq e^{\epsilon} \Prob(M(d') \in S) + \delta, \;\; \forall S \in \mathcal M. 
\end{align}
If $\delta=0$, the mechanism is said to be $\epsilon$-differentially private. 
\end{definition}
This definition quantifies the admissible deviations for the output distribution of a differentially private mechanism, 
when a variation at the input satisfies the adjacency relation.
Smaller values of $\epsilon$ and $\delta$ correspond to stronger privacy guarantees.
In this paper, the space $\D$ was defined as the space of input signals $y$, the adjacency relation considered 
is \eqref{eq: adjacency relation 0} or \eqref{eq: adjacency relation}, and the output space $\R$ is the space of output 
signals for the observer, here $\mathsf X^{\mathbb N}$ since we wish to estimate $x$.  
The problem is to publish an accurate estimate of the state $x$ while satisfying the property of 
Definition \ref{def: differential privacy original} for specified values of $\epsilon$ and $\delta$.

\begin{remark}
Definition \ref{def: differential privacy original} depends on the choice of $\sigma$-algebra $\mathcal M$,
which must contain enough sets $S$ to provide a meaningful differential privacy guarantee. 
The interested reader can find a discussion of measurability issues in a previous paper\cite{LeNyDP2012_journalVersion:1}.
\end{remark}

\subsection{Sensitivity and Basic Differentially Private Mechanisms}		\label{section: sensitivity}

Enforcing differential privacy can be done by randomly perturbing the published output of a system 
\cite{Dwork06_DPcalibration:1, LeNyDP2012_journalVersion:1}, at the expense of its quality or utility.
Hence, we are interested in evaluating as precisely as possible the amount of noise necessary 
to make a mechanism differentially private. For this purpose, the following quantity plays an important role.

\begin{definition}	\label{def: sensitivity}
Let $q \geq 1$. The $\ell_q$-sensitivity of a system $G$ with $m$ inputs and $n$ outputs with respect to an 
adjacency relation $\Adj$ is defined by
$\Delta_q G = \sup_{\text{Adj(u,u')}} \| Gu - Gu' \|_q$.
\end{definition} 
In practice we are interested in the sensitivity of a system for the cases $q=1$ and $q=2$.
The basic mechanisms of Theorem \ref{eq: basic DP mechanism} below 
(with proofs and references in a previous paper\cite{LeNyDP2012_journalVersion:1}), 
can be used to produce differentially private signals.
First, we need the following definitions. A zero-mean Laplace random variable with parameter $b$
has the probability density function $\exp(-|x|/b)/2b$, and its variance is $2 b^2$. The $\mathcal Q$-function is defined as
$\mathcal Q(x) := \frac{1}{\sqrt{2 \pi}} \int_x^{\infty} e^{-\frac{u^2}{2}} du$. 
Then, for $\epsilon > 0$, $0.5 \geq \delta > 0$,
let $K = \mathcal Q^{-1}(\delta)$ and define $\kappa_{\delta,\epsilon} = \frac{1}{2 \epsilon} (K+\sqrt{K^2+2\epsilon})$,
which can be shown to behave roughly as $O\left( \sqrt{\ln(1/\delta)}/\epsilon \right)$.

\begin{theorem}	\label{eq: basic DP mechanism}
Let $G$ be a system with $m$ inputs and $n$ outputs, and fix a relation $\text{Adj}$ in Definition \ref{def: sensitivity}.
The mechanism $Mu = Gu + w$, where all $w_{k,i}, k \in \mathbb N, 1 \leq i \leq n$, are independent
Laplace random variables with parameter $b \geq (\Delta_1 G)/\epsilon$, is  $\epsilon$-differentially private 
for $\Adj$.
If $w$ is instead a white Gaussian noise such that the covariance matrix of each sample $w_k$ is $\sigma^2 I_n$ with 
$\sigma \geq \kappa_{\delta,\epsilon} \, \Delta_2 G$, then the mechanism is $(\epsilon,\delta)$-differentially private.
\end{theorem}

The mechanisms of Theorem \ref{eq: basic DP mechanism} are called the Laplace and the Gaussian mechanism. 
One reason for introducing the Gaussian mechanism is that typically the $\ell_2$-sensitivity is smaller than its
$\ell_1$ counterpart, which leads to lower noise levels if one can tolerate $\delta > 0$ in the privacy guarantee.

\subsection{Input and Output Perturbation}

\begin{figure}
\centering
\includegraphics[width=0.6\linewidth]{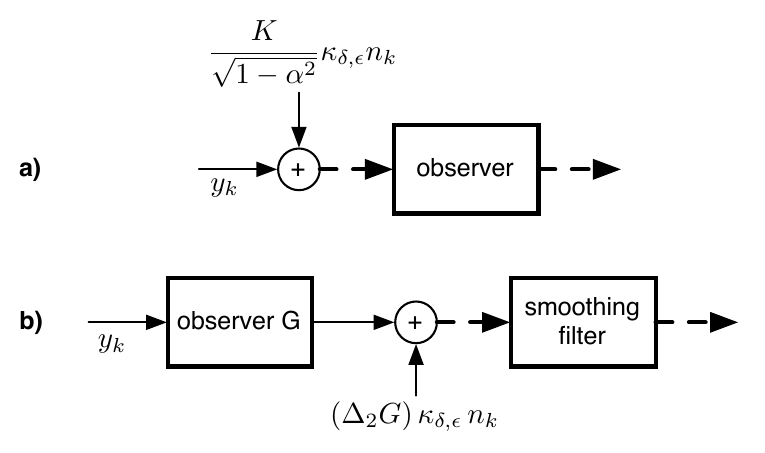}
\caption{Gaussian mechanisms with input (a) and output (b) perturbation for the adjacency relation \eqref{eq: adjacency relation}. 
$n_k$ represents a zero-mean standard white Gaussian noise with identity covariance matrix.
Dashed lines represent a differentially private signal.}
\label{fig: input and output perturbation}
\end{figure}

Theorem \ref{eq: basic DP mechanism} says that we can obtain a differentially private signal
at the output of a system $G$ by adding noise with standard deviation  
proportional to $\Delta_1G/\epsilon$ or to $\kappa_{\delta,\epsilon} \Delta_2 G$.
A very useful additional result stated here informally says that post-processing a differentially private signal
without re-accessing the privacy-sensitive input signal does not change the differential privacy guarantee
\cite{LeNyDP2012_journalVersion:1}.
Now, the system $G$ in Theorem \ref{eq: basic DP mechanism} can simply be the identity, with
$\ell_1$- and $\ell_2$- sensitivity for the adjacency relation \eqref{eq: adjacency relation}  
equal to $K/(1-\alpha)$ and $K/\sqrt{1-\alpha^2}$ 
respectively (and $B_1$ and $B_2$ for \eqref{eq: adjacency relation 0}). 
This immediately gives a first possible design approach for our privacy-preserving observer, 
simply adding Laplace or Gaussian noise directly to the input signal $y$, 
see Fig. \ref{fig: input and output perturbation} a). The observer can then be designed according
to any desired methodology, and should try to mitigate the effect of the artificial input noise, whose 
distribution is known, in addition to the usual measurement error. We call this design an input perturbation 
mechanism. Note that for $\alpha$ close to $1$, $1/\sqrt{1-\alpha^2}$ is significantly smaller   
than $1/(1-\alpha)$, so that if we are willing to accept some $\delta > 0$ in the privacy guarantee and to
use the $2$-norm on $\mathsf Y$ in the adjacency relation \eqref{eq: adjacency relation},  
we can obtain much better accuracy by using the $\ell_2$-sensitivity.

The input perturbation mechanism is attractive for its simplicity and might perform well,
especially with low privacy level requirements (high $\epsilon$, $\delta$). 
In particular, the sensitive data can be made differentially private at the source,
before sending it to any third party for processing. 
However, it can also potentially exhibit the following drawbacks. 
First, the noise added to $y$ might be unnecessarily large because it is 
not tailored to the task of estimating the state $x$ of the model 
\eqref{eq: model - 1}-\eqref{eq: model - 2}, and does not take into account 
the temporal correlations between samples of the signal $y$ captured by this model. 
Significant noise at the input of the observer can also lead to poor performance, i.e., slow 
convergence and large errors in the state estimate, or even perhaps divergence of the
estimate from the true state trajectory, since the convergence of nonlinear observers is often local.
Second, characterizing the output error (state estimation error) due to the privacy-preserving noise requires 
understanding how this noise is transformed when passing through the nonlinear observer. 
In general, for nonlinear systems, the noise distribution at the output can become multimodal and 
non-zero mean, and hence the observer could produce a systematically biased estimate that
could be hard to correct.

An alternative to input perturbation is the output perturbation mechanism shown on Fig. \ref{fig: input and output perturbation} b).
In this case, following Theorem \ref{eq: basic DP mechanism}, a privacy-preserving noise signal proportional to the sensitivity 
of the observer $G$ is added at its output. 
Computing the sensitivity of $G$, or in practice upper bounding it, should be done as accurately as possible to reduce 
the conservatism of the approach. 
On the other hand the output noise does not impact any stability or bias analysis of the observer $G$. 
As discussed in more details in the following sections, we should then try to design an observer 
that has both good tracking performance for the state trajectory and low sensitivity, in order 
to minimize the level of privacy-preserving noise necessary at the output. 
These two desired properties are essentially in conflict.
Fig. \ref{fig: input and output perturbation} b) shows that we can also add a terminal filter 
to smooth out the Laplace or Gaussian noise\cite{Cortes:DPtuto:CDC16}, although this can generally affect 
the transient performance of the overall system (e.g., its convergence speed). 
We do not discuss the design of a potential smoothing filter in this paper, except briefly in Subsection \ref{section: DSBM}.

\begin{eexample}
Consider the memoryless system $y_k \mapsto \phi(y_k) := y_k^2$,
which could be a simple state estimator for a measurement model
$y_k = \sqrt{x_k}$ in  \eqref{eq: model - 2}, 
which does not take the dynamics \eqref{eq: model - 1} into account.
Consider the adjacency relation \eqref{eq: adjacency relation} for $\alpha=0$, 
so that we have a deviation at some (unknown) single time period $k_0$ of 
at most $K$ between adjacent signals $y_k$ and $\tilde y_k$. 
For the input perturbation scheme and the Gaussian mechanism, assuming for simplicity 
that $\kappa_{\delta,\epsilon} = 1$, the signal 
$
z_k = (y_k+K \xi_k)^2 = y_k^2 + 2 K y_k \xi_k + K^2 \xi_k^2 
$
is differentially private when $\xi$ is a standard Gaussian white noise. 
The privacy-preserving noise at the input induces a systematic bias at
the output between $z_k$ and $y_k^2$ equal to 
$\mathbb E[2 K y_k \xi_k + K^2 \xi_k^2] = K^2$.
Since $K$ is assumed publicly known, in this case the bias can be compensated and 
a better approximation of $\phi$ that is still differentially private is $z_k' = (y_k + K \xi_k)^2 - K^2$.
One can verify that the variance of the remaining error is $e'_k = \mathbb E[(z_k'-y_k^2)^2] = 4 K^2 y_k^2 + 2 K^4$.

Suppose we know in addition that $y_k \in [0,1]$ for all $k \geq 0$.
Then we can bound the sensitivity of the memoryless system as 
\begin{equation}	\label{eq: memoryless example}
\Delta_2\phi = |y_{k_0}^2 - \tilde{y}_{k_0}^2| 
= |y_{k_0} - \tilde{y}_{k_0}| |y_{k_0} + \tilde{y}_{k_0}| 
\leq 2 |y_{k_0} - \tilde{y}_{k_0}| \leq 2 K.
\end{equation}
Hence, the signal $z_k'' = y_k^2 + 2 K \xi_k$ is also differentially private and unbiased, 
with $\xi$ a standard white Gaussian noise as before. The variance of the error is 
$e_k'' = \mathbb E[(z_k''-y_k^2)^2] = 4 K^2$, which is smaller than the worst case value
$4K^2 + 2K^4$ for $e_k'$. However, $e_k''$ is larger than $e_k'$ as soon as $y_k < \sqrt{1-K^2/2}$,
the typical case since $K$ should be much less than $1$, otherwise both the input and output mechanisms
essentially destroy the signal. 
The upper bound \eqref{eq: memoryless example} on the sensitivity is conservative 
in order to be independent of the actual values of the sensitive signal $y$, which is necessary when
Theorem \ref{eq: basic DP mechanism} is used to provide a differential privacy guarantee.
\end{eexample}

In the rest of the paper we focus on the output perturbation mechanism of Fig. \ref{fig: input and output perturbation} b).
There are two aspects to the differentially private observer design problem in this case. First, we need to enforce 
appropriate convergence of $z$ toward $x$, which is the observer design problem itself. Second, we
also need to control and bound explicitly the magnitude of the changes in $z$ when the observer input changes 
from $y$ to an adjacent signal $\tilde y$, in order to apply Theorem \ref{eq: basic DP mechanism} and set
the output noise level providing the differential privacy guarantee.
In this paper, both aspects of the problem are treated by using contraction analysis to design the observer 
as well as quantify its sensitivity to variations in the measured signal $y$.
A motivation for this approach is the exponential convergence of trajectories of contractive systems
toward each other, which provides a degree of robustness against input disturbances 
\cite{Lohmiller98_contraction:1, Sontag10_contraction:1, Chung_sync13:1,KhalilBook02:1}
and, as a consequence, sensitivity bounds for variations in input data streams $y$.
The next section provides some background on contraction analysis, 
necessary to describe our design approach in Section \ref{section: DP observers}.


\section{Contracting Systems}		\label{section: contraction}

Contraction analysis is an ``incremental'' stability analysis methodology for dynamical systems 
emphasizing convergence of trajectories toward each other, popularized in particular by 
the work of Lohmiller and Slotine \cite{Lohmiller98_contraction:1}. Earlier related work can also 
be found in the mathematics literature \cite{Lewis49:incrementalStability:1, Hartman82:ODEbook}.
Contraction and incremental stability analysis have seen significant developments in 
the past two decades \cite{Lohmiller98_contraction:1, Angeli00_contraction:1, Aghannan:TAC03:contractingObs,
Pavlog06:contractionBook:1, Chung_sync13:1, Russo_contraction_PLOS10:1, Sontag10_contraction:1, Forni14_contraction:1},
and we refer the reader to the recent paper by Forni and Sepulchre \cite{Forni14_contraction:1} for
a comparison of different variations that have emerged and additional references.
The purpose of this section is to review some aspects of this methodology for discrete-time systems,
which are not emphasized as much as continuous-time systems in the literature, and to state and 
prove some results that we rely on to design differentially private observers with output perturbation. 
Although these results could potentially be derived from the ideas presented in some of the references   
cited above, we provide here a self-contained discussion and in particular explicit bounds on distances
between trajectories that are necessary to precisely set the level of privacy-preserving noise, 
since qualitative guarantees of incremental convergence are insufficient.

\subsection{Basic Results}	\label{section: basic contraction}

Consider a discrete-time system
\begin{equation}	\label{eq: DT system}
x_{k+1} = f_k(x_k),
\end{equation}
with $f_k: \mathsf X \to \mathsf X$ a $\mathcal C^1$ function, for all $k \in \mathbb N$.  
Let us denote by $\phi(k;k_0,x_0)$ the value at time $k \geq k_0$ of the solution 
of (\ref{eq: DT system}) taking the value $x_0$ at time $k_0$.
A forward invariant set for the system (\ref{eq: DT system}) is a set $C \subset \mathbb \mathsf X$ 
such that if $x_0 \in C$, then for all $k_0$ and all $k \geq k_0$, $\phi(k;k_0,x_0) \in C$. 
Although we assume in this paper $\mathsf X = \mathbb R^n$, it is useful to introduce here some language
from differential geometry and view $\mathsf X$ more generally as an $n$-dimensional differentiable 
manifold\cite{doCarmo:RiemannianGeomBook, Forni14_contraction:1}. 
For each point $x \in \mathsf X$, the tangent space to $\mathsf X$ at $x$, i.e., informally, the $n$-dimensional 
vector space of all tangent vectors to curves on $\mathsf X$ passing through $x$, is denoted $\mathsf T_x \mathsf X$.
The tangent bundle of $\mathsf X$ is denoted $\mathsf T \mathsf X := \cup_{x \in \mathsf X} \{x\} \times \mathsf T_x \mathsf X$,
and is equipped with a time-varying family of norms $|\cdot|_{[x,k]}$, smoothly varying with $x$, so that 
$|\cdot|_{[x,k]}$ is a norm on $\mathsf T_x \mathsf X$, for all $k \in \mathbb N$.
For each $x, \tilde x \in \mathsf X$, let $\Gamma(x,\tilde x)$ be the set of piecewise $\mathcal C^1$ curves
joining $x$ and $\tilde x$, i.e., functions $\gamma: [0,1] \to \mathsf X$ with $\gamma(0) = x$, $\gamma(1) = \tilde x$.
We define the (time-varying) length of such a curve $\gamma$ by
\[
L_k(\gamma) = \int_0^1 |\gamma'(r)|_{[\gamma(r),k]} dr,
\]
where $\gamma'(r) := \frac{d\gamma}{dr}(r)$. We then have a notion of (time-varying) geodesic distance on $\mathsf X$, defined as
\begin{align}	\label{eq: geodesic distance}
d_k(x, \tilde x) = \inf_{\gamma \in \Gamma(x,\tilde x)} L_k(\gamma), \; \forall x, \tilde x \in \mathsf X.
\end{align}
Moreover, if the norms $|\cdot|_{[x,k]}$ are in fact independent of $x$, thus denoted $|\cdot|_{[k]}$, 
and if $\mathsf X$ is a convex set in $\mathbb R^n$ (possibly equal to $\mathbb R^n)$, then the infimum 
in \eqref{eq: geodesic distance} is achieved by straight lines $\gamma(r) = x + r (\tilde x - x)$ 
and $d_k(x, \tilde x) = |\tilde x - x|_{[k]}$ in \eqref{eq: geodesic distance}.
Finally, each function $f_k$ in \eqref{eq: DT system} is associated to a Jacobian
$F_k(x) := \frac{\partial f_k}{\partial x}(x)$, which defines a linear map from $\mathsf T_x \mathsf X$ at
time $k$ to $\mathsf T_{f_k(x)} X$ at time $k+1$. 
As a result, for all vectors $v \in \mathsf T_x \mathsf X$,   
\begin{align}	\label{eq: induced norm def. inequality}
|F_k(x) \, v|_{[f_k(x),k+1]} \leq \|F_k(x)\|_{[x.k]}^{[f_k(x),k+1]} \, |v|_{[x,k]},
\end{align}
where $\| \cdot \|_{[x.k]}^{[f(x),k+1]}$ denotes the norm induced by $|\cdot|_{[x,k]}$ and $|\cdot|_{[f_k(x),k+1]}$.

\begin{remark}
The discussion could be carried out in a slightly more general framework by allowing 
asymmetric norms on the tangent spaces\cite{Forni14_contraction:1} rather than standard
norms, but we will not need this level of generality.
\end{remark}

\begin{definition}	\label{def: contractivity}
Let $\rho$ be a nonnegative constant.
The system (\ref{eq: DT system}) is said to be \emph{$\rho$-contracting} for the 
norms $| \cdot |_{[x,k]}$ on a forward invariant set $C \subset \mathsf X$ if 
for any $k_0 \in \mathbb N$ and any two initial conditions $x_{0}, \tilde x_{0} \in C$, 
we have, for all $k \geq k_0$,
\begin{align}	\label{def: geom contractivity}
d_k(\phi(k;k_0,x_0), \phi(k;k_0,\tilde x_0)) \leq \rho^{k-k_0} d_{k_0}(x_{0}, \tilde x_{0}).
\end{align}
\end{definition}

Let $\gamma_k \in \Gamma(x, \tilde x)$ be a curve joining two points $x$ and $\tilde x$ in $\mathsf X$ at a fixed time $k$.
Let $\gamma_k'(r)$ be the tangent vector to $\gamma_k$ at the point $\gamma_k(r)$, for $r \in [0,1]$.
The curve $\gamma_k$ is transported at time $k$ by \eqref{eq: DT system} to a curve $\gamma_{k+1}$ joining $f_k(x)$ and $f_k(\tilde x)$.
Taking the derivative with respect to $r$ in the equation $\gamma_{k+1}(r) = f_k(\gamma_k(r))$, we obtain the important
\emph{linear} relation between tangent vectors
\begin{align}	\label{eq: linear relation tangent vectors}
\gamma_{k+1}'(r) = F_k(\gamma_k(r)) \, \gamma_k'(r), \; \forall r \in [0,1], \forall k \geq 0.
\end{align}
The following fundamental theorem of contraction analysis is then a consequence of \eqref{eq: linear relation tangent vectors}.

\begin{theorem}		\label{thm: contractivity condition simple}
Let $F_k = \frac{\partial f_k}{\partial x}$ be the Jacobian of $f_k$, for all $k \geq 0$. 
A sufficient condition for the system \eqref{eq: DT system} to be $\rho$-contracting 
for the norms $| \cdot |_{[x,k]}$ on a forward invariant set $C \subset \mathsf X$
is that 
\begin{align}	\label{eq: contraction sufficient condition}
\|F_k(x)\|_{[x.k]}^{[f_k(x),k+1]} \leq \rho, \;\; \forall x \in C, \forall k \in \mathbb N.
\end{align}
\end{theorem}

\begin{proof}
Consider a curve $\gamma_{k_0}: [0,1] \to \mathsf X$ in $\Gamma(x_0,\tilde x_0)$.
This curve is transported by \eqref{eq: DT system} to a sequence of curves $\gamma_{k_0+1}, \gamma_{k_0+2}, \ldots$, 
i.e., $\gamma_{k+1}(r) = f_k(\gamma_k(r))$, for all $r \in [0,1]$, with $\gamma_k$ joining
$\phi(k;k_0,x_0)$ and $\phi(k;k_0,\tilde x_0)$. 
We have, for all $k \geq k_0$, using  \eqref{eq: linear relation tangent vectors}
\begin{align}
L_{k+1}(\gamma_{k+1}) = \int_0^1 |\gamma_{k+1}'(r)|_{[\gamma_{k+1}(r),{k+1}]} dr = 
\int_0^1 |F_k(\gamma_k(r)) \, \gamma_k'(r)|_{[\gamma_{k+1}(r),k+1]} dr. \nonumber 
\end{align}
Now, using \eqref{eq: induced norm def. inequality} and then the assumption \eqref{eq: contraction sufficient condition}
\begin{align}
L_{k+1}(\gamma_{k+1}) &\leq \int_0^1 \|F_k(\gamma_k(r))\|_{[\gamma_k(r),k]}^{[\gamma_{k+1}(r),k+1]} \, |\gamma_{k}'(r)|_{[\gamma_{k}(r),k]} dr 
\leq \rho \int_0^1 |\gamma_{k}'(r)|_{[\gamma_{k}(r),k]} dr = \rho L_k(\gamma_k), \label{eq: induced norm inequality}
\end{align} 
and hence by immediate recursion, $L_k(\gamma_k) \leq \rho^{k-k_0} L_{k_0}(\gamma_{k_0})$. To conclude,
let $\epsilon > 0$ and take the curve $\gamma_{k_0}$ above to satisfy 
\[
L_{k_0}(\gamma_{k_0}) \leq (1+\epsilon) d_{k_0}(x_0, \tilde x_0). 
\]
Then, since $\gamma_k \in \Gamma(\phi(k;k_0,x_0), \phi(k;k_0,\tilde x_0))$, we have
\begin{equation}	\label{eq: stronger contraction inequality}
d_k(\phi(k;k_0,x_0), \phi(k;k_0,\tilde x_0)) \leq L_k(\gamma_k) \leq \rho^{k-k_0} L_{k_0}(\gamma_{k_0}) 
\leq (1+\epsilon) \rho^{k-k_0} d_{k_0}(x_0, \tilde x_0).
\end{equation}
Since this inequality is true for all $\epsilon > 0$, \eqref{def: geom contractivity} holds.
\end{proof}

\begin{remark}
Note that to obtain useful results in continuous time (in particular, to detect convergent dynamics), it is crucial
to use a tighter inequality replacing the first inequality of \eqref{eq: induced norm inequality} by Coppel's   
inequality \cite{Vidyasagar:book93:nonlinear} to bound the solutions of linear differential equations. 
This leads to a sufficient condition for continuous-time systems similar to \eqref{eq: contraction sufficient condition} stated
in terms of matrix measures instead of induced norms \cite{Lohmiller98_contraction:1,Russo_contraction_PLOS10:1,Sontag10_contraction:1}. 
However, this does not apply to discrete-time systems.
\end{remark}

\begin{corollary}
With the notation defined as in Theorem \ref{thm: contractivity condition simple}, suppose that 
$C$ is a convex forward invariant subset of $\mathbb R^n$
and that the norms $| \cdot |_{[x,k]}$  on the tangent spaces are independent of $x$  
and denoted $| \cdot |_k$. Let $\| \cdot \|_k^{k+1}$ be the matrix norm induced by $| \cdot |_k$ and $| \cdot |_{k+1}$. 
Then, if $\|F_k(x)\|_k^{k+1} \leq \rho$ for all $x \in C$ and for all $k \in \mathbb N$, we have
\[
|\phi(k;k_0,x_0) - \phi(k;k_0,\tilde x_0)|_k \leq \rho^{k-k_0} |x_{0} - \tilde x_{0}|_{k_0}, \;\; \forall x_0, \tilde x_0 \in C, \forall k \geq k_0.
\]
\end{corollary}
\begin{proof}
The result follows immediately from Theorem \ref{thm: contractivity condition simple} and the remarks on 
geodesic distances preceding Definition \ref{def: contractivity}.
\end{proof}

\begin{corollary}	\label{cor: contraction result 1-norm}
With the notation defined as in 
Theorem \ref{thm: contractivity condition simple}, suppose that the norms on the tangent spaces 
are defined for all $x$ and $k$ by $| v |_{[x,k]} = |P_{[x,k]} v|_1$, where $P_{[x,k]} = \text{diag}(p_{[x,k]})$, 
with $p_{[x,k]}$ a vector with positive components $p_{[x,k],i}$. Hence, $| v |_{[x,k]} = \sum_{i=1}^n p_{[x,k],i} |v_i|$.
Then the system is $\rho$-contracting for the associated distances on $\mathsf X$ if the following linear programs
are feasible, for all $x \in C$ and $k \in \mathbb N$
\begin{align}	
&\sum_{i=1}^n p_{[f_k(x),k+1],i} |F_{k,ij}(x)| \leq \rho \, p_{[x,k],j}, \;\; \forall 1 \leq j \leq n, \label{eq: contraction l1 condition - 1} \\
&p_{[x,k], i}, p_{[f_k(x),k+1], i} > 0, \; \forall 1 \leq i \leq n.  \label{eq: contraction l1 condition - 2}
\end{align}
In particular, if $C$ is convex and if there exist positive vectors $p_{[k]}$ independent of $x$ satisfying the above
inequalities \eqref{eq: contraction l1 condition - 1}, \eqref{eq: contraction l1 condition - 2} 
for all $x, k$, then, with $P_{[k]} := \text{diag}(p_{[k]})$, $x_k := \phi(k;k_0,x_0)$, $\tilde x_k := \phi(k;k_0,\tilde x_0))$, 
we have
\begin{align}	\label{eq: special case constant matrices - l1}
|P_{[k]} (x_k - \tilde x_k) |_1 \leq \rho^{k-k_0} |P_{[k_0]}(x_0 - \tilde x_0)|_1, \;\; \forall x_0, \tilde x_0 \in C, \forall k \geq k_0.
\end{align}
\end{corollary}

\begin{proof}
The inequalities \eqref{eq: contraction l1 condition - 1}, \eqref{eq: contraction l1 condition - 2} come from
satisfying \eqref{eq: contraction sufficient condition} for the $1$-norms weighted by $R := P_{[x,k]}$ and $S := P_{[f_k(x),k+1]}$.
The condition \eqref{eq: contraction sufficient condition} is equivalent to the induced $1$-norm of the matrix
$S F_k(x) R^{-1}$ being less than $\rho$, and this matrix has entries 
$p_{[f_k(x),k+1], i} F_{k,ij}(x) / p_{[x,k],j}$.
The induced $1$-norm of an $n \times m$ matrix $A = [a_{ij}]_{i,j}$ 
is $\max_{1 \leq j \leq m} \sum_{i=1}^n |a_{ij}|$. The result follows from these facts.
\end{proof}

\begin{corollary}	\label{cor: classical contraction theorem}
With the notation defined as in Theorem \ref{thm: contractivity condition simple}, suppose that the norms on the
tangent spaces are defined by $| v |_{[x,k]} = (v^T P_{[x,k]} v)^{1/2} = |P_{[x,k]}^{1/2} v|_2$, where $P_{[x,k]} \succ 0$, for all $x$ and $k$.
Then the system is $\rho$-contracting for the associated distances on $\mathsf X$ if
the following Linear Matrix Inequalities (LMIs) are satisfied
\begin{equation}	\label{eq: contraction l2 condition}
F_k(x)^T P_{[f(x),k+1]} F_k(x) \preceq \rho^2 P_{[x,k]}, \; \forall x \in C, \forall k \in \mathbb N.
\end{equation}
Suppose $C$ is convex.
If there exist matrices $P_{[k]} \succ 0$, $k \in \mathbb N$, independent of $x$, satisfying these LMIs, 
then we have
\begin{equation}	\label{eq: special case constant matrices - l2}
|P_{[k]}^{1/2} (x_k - \tilde x_k)|_2 \leq \rho^{k-k_0} |P_{[k_0]}^{1/2} (x_0 - \tilde x_0)|_2, \;\; \forall x_0, \tilde x_0 \in C, \forall k \geq k_0,
\end{equation}
where $x_k := \phi(k;k_0,x_0)$, $\tilde x_k := \phi(k;k_0,\tilde x_0))$.
If there exist matrices $P_{[x,k]}$ satisfying \eqref{eq: contraction l2 condition} and if
there there exist $2$ matrices $P_{\min} \succ 0$ with minimum eigenvalue $\lambda_{\min} > 0$ 
and $P_{\max} \succ 0$ with maximum eigenvalue $\lambda_{\max} > 0$ such that
we have $\lambda_{min} I \preceq P_{\min} \preceq P_{[x,k]} \preceq P_{\max} \preceq \lambda_{max} I$,
for all $x, k$, then 
\[
|P_{\min}^{1/2}(x_k - \tilde x_k)|_2 \leq \rho^{k-k_0}  |P_{\max}^{1/2}(x_0 - \tilde x_0)|_2,  \;\; \forall x_0, \tilde x_0 \in C, \forall k \geq k_0,
\]
and hence
\[
|x_k - \tilde x_k|_2 \leq \rho^{k-k_0} \sqrt{\frac{\lambda_{\max}}{\lambda_{\min}}} |x_0- \tilde x_0|_2.
\]
\end{corollary}
\begin{proof}
This is a corollary of Theorem \ref{thm: contractivity condition simple}, since satisfying \eqref{eq: contraction sufficient condition}
for the norm induced by the weighted $2$-norms with matrices $P_{[x,k]}$ and $P_{[f(x),k+1]}$ can be written
$v^T F_k(x)^T P_{[f(x),k+1]} F_k(x) v \leq \rho^2 \, v^T P_{[x,k]} v$, for all $v$ in $\mathbb R^n$.
The second part uses the fact
\[
\int_0^1 \sqrt{\gamma'(r) P_{\max} \gamma'(r)} dr \geq L_k(\gamma) = \int_0^1 \sqrt{\gamma'(r) P_{[\gamma(r),k]} \gamma'(r)} dr \geq 
\int_0^1 \sqrt{\gamma'(r) P_{\min} \gamma'(r)} dr,
\]
and moreover $\int_0^1 \sqrt{\gamma'(r) P_{\min} \gamma'(r)} dr \geq |P_{min}^{1/2} (x - \tilde x)|_2$ if $\gamma \in \Gamma(x,\tilde x)$,
since for a constant norm on $\mathbb R^n$ the geodesic curves are straight lines. 
Finally, referring to \eqref{eq: stronger contraction inequality}, we get
\[
|P_{min}^{1/2} (x_k - \tilde x_k)|_2 \leq L_k(\gamma_k) \leq \rho^{k-k_0} L_{k_0}(\gamma_{k_0}) \leq \rho^{k-k_0} |P_{\max}^{1/2}(x_0 - \tilde x_0)|_2.
\]
\end{proof}

\begin{remark}
Corollary \ref{cor: classical contraction theorem} is the classical contraction result \cite{Lohmiller98_contraction:1}, 
in discrete time, for norms associated with an inner product (Riemannian structure on $\mathsf X$).
Using state-dependent $P$ matrices can enlarge the set of systems for which we can prove
contraction, but in our case we also need to explicitly 
bound the Euclidean distances $|x_k - \tilde x_k|_2$, not just general geodesic distances, 
to be able to evaluate the level of noise necessary for the Gaussian mechanism of Theorem \ref{eq: basic DP mechanism}.
\end{remark}

\subsection{Effect of Disturbances}	\label{section: disturbance effects}

For the computation of $\ell^1$ and $\ell^2$-sensitivities, we need to study the 
trajectory deviations of contracting system subject to disturbances.
Qualitatively, the exponential convergence of trajectories of a contracting system 
provides some robustness against disturbances\cite{KhalilBook02:1,Lohmiller98_contraction:1,Sontag10_contraction:1}.
However, to precisely set the level of privacy-preserving noise, quantitative worst case bounds 
on the $\ell^1$ or $\ell^2$-norms of the trajectory deviations are needed.  
Hence, consider a system
\begin{equation}	\label{eq: perturbed contractive system with disturbance}
x_{k+1} = f_k(x_k, \pi_k(x_k)),
\end{equation}
where $\pi_k: \mathsf X \to \mathsf{P} := \mathbb R^p$, for some $p$, represents a $\mathcal C^1$ disturbance signal, and
for all $k \geq 0$, $f_k: \mathsf X \times \mathsf P \to \mathsf X$ is $\mathcal C^1$. 
We equip the tangent spaces of the product manifold $\mathsf X \times \mathsf{P}$ with time-varying norms that for 
simplicity are assumed to be fixed for the disturbance part, i.e., $| (v,w) |_{[(x,\pi),k]} = |v|_{[x,k]} + |w|_\mathsf{P}$, 
for a fixed norm $|\cdot|_\mathsf{P}$. The nominal system under zero disturbance is 
\begin{equation}	\label{eq: nominal system}
\bar x_{k+1} = f_k(\bar x_k, 0).
\end{equation}
We denote $\frac{\partial f_k}{\partial x}$ and $\frac{\partial f_k}{\partial \pi}$ the Jacobian of $f_k(x,\pi)$ with respect
to the components of $x$ and $\pi$ respectively.
For $r \in [0,1]$, denote by $\phi(k;r,k_0,x_0)$ the iterates of 
\begin{equation}	\label{eq: perturbed system with r}
x_{k+1} = f_k(x_k, r \, \pi_k(x_k)),
\end{equation}
starting from $x_0$ at time $k_0$. Note that \eqref{eq: perturbed contractive system with disturbance} corresponds
to $r = 1$ and \eqref{eq: nominal system} to $r = 0$.
Let us also define
\begin{align}  	\label{eq: Jacobian parametrized}
J_k(x;r) &:= \frac{\partial f_k}{\partial x}(x,r \, \pi_k(x)) + r \frac{\partial f_k}{\partial \pi}(x,r \, \pi_k(x)) \frac{\partial \pi_k}{\partial x}(x),
\; \forall x \in \mathsf X, \forall r \in [0,1]. 
\end{align}
For all $x$ in $\mathsf X$, denote $x_+^{k,r} := f_k(x,r \, \pi_k(x))$.
Formally, the ``differential'' maps \eqref{eq: Jacobian parametrized} are from $\mathsf T_{[x,k]} \mathsf X$ to 
$\mathsf T_{[x_+^{k,r},k+1]} \mathsf X$,
with the corresponding induced norms 
$\| \cdot \|_{[x,k]}^{[x_+^{k,r},k+1]}$.
We then have the following result. 

\begin{theorem}	\label{thm: contractivity under disturbance}
Consider a trajectory 
$\bar x_k := \phi(k;0,k_0,\bar x_0)$ for \eqref{eq: nominal system} starting from $\bar x_0$ and a
trajectory 
$x_k := \phi(k;1,k_0,x_0)$ for the perturbed system \eqref{eq: perturbed contractive system with disturbance} 
starting from $x_0$.
We suppose that there exists a sequence $\{M_k\}_{k \geq 0}$ such that 
\begin{equation}	\label{eq: bound on disturbance manifold}
\left | \frac{\partial f_k}{\partial \pi}(x, r \, \pi_k(x)) \, \pi_k(x) \right|_{[x_+^{k,r},k+1]} \leq M_k, 
\;\;  \forall r \in [0,1], \forall x \in C, \forall k \geq k_0,
\end{equation}
and that
\begin{equation} \label{eq: contraction condition perturbed 1}
\| J_k(x;r) \|_{[x,k]}^{[x_+^{k,r},k+1]} \leq \rho, \;\;  \;\; \forall r \in [0,1], \forall x \in C, \forall k \geq k_0,
\end{equation}
where $C$ is a forward invariant set for 
\eqref{eq: perturbed system with r}, for all $r \in [0,1]$.
Then we have, for all $k \geq k_0$, and the distances $d_k$ defined in \eqref{eq: geodesic distance},
\[
d_k(\bar x_k, x_k) \leq \rho^{k-k_0} d_{k_0}(\bar x_{k_0}, x_{k_0}) + \sum_{l=0}^{k-k_0-1} \rho^l M_{k-1-l}.
\]
\end{theorem}

\begin{remark}
As an example, in the case of additive disturbances on $\mathsf X = \mathsf P = \mathbb R^n$, i.e., 
\begin{equation}	\label{eq: additive disturbance}
f_k(x, \pi_k(x)) = \tilde f_k(x) + \pi_k(x),
\end{equation}
with a fixed norm $|\cdot|$ on $\mathbb R^n$, the condition \eqref{eq: bound on disturbance manifold} can be written 
more simply $\sup_{x \in \mathsf C} |\pi_k(x)| \leq M_k$. 
\end{remark}

\begin{remark}	\label{rem: simplification of thm}
Note that if the disturbance $\pi_k$ does not depend on $x$, then \eqref{eq: Jacobian parametrized}
reads $J_k(x;r) := \frac{\partial f_k}{\partial x}(x,r \, \pi_k)$ and \eqref{eq: contraction condition perturbed 1} is
a type of contraction condition on the perturbed system. If moreover the perturbation is in fact additive as in
\eqref{eq: additive disturbance}, then $\eqref{eq: contraction condition perturbed 1}$ simply asks that the
Jacobian of the nominal system $\tilde f_k$ satisfy the contraction assumption. 
\end{remark}

\begin{proof}
Consider a curve $\gamma_{k_0} \in \Gamma(\bar x_{0}, x_{0})$, i.e., such that $\gamma_{k_0}(0) = \bar x_{0}$ 
and $\gamma_{k_0}(1) = x_{0}$, 
transported by \eqref{eq: perturbed system with r} to the sequence 
\[
\gamma_{k}(r) = \phi(k; r, k_0, \gamma_{k_0}(r)), \; \forall r \in [0,1], \forall k \geq k_0.   
\]
Then, for $k \geq k_0$, we have $\gamma_k \in \Gamma(\bar x_k, x_k)$, where $\bar x_k := \phi(k;0,k_0,\bar x_0)$ and
$x_k := \phi(k;1,k_0,x_0)$. 
Following the idea of the proof of Theorem \ref{thm: contractivity condition simple}, define 
$\gamma_k'(r) := \frac{d}{dr} \phi(k;r,k_0,\gamma_{k_0}(r))$,
so that we have, for all $k$ and all $r \in [0,1]$
\[
\gamma'_{k+1}(r) = J_k(\gamma_k(r);r) \, \gamma'_k(r) + \frac{\partial f_k}{\partial \pi}(\gamma_k(r), r \, \pi_k(\gamma_k(r))) \, \pi_k(\gamma_k(r)), 
\]
which implies, by \eqref{eq: contraction condition perturbed 1} and \eqref{eq: bound on disturbance manifold},
\[
|\gamma'_{k+1}(r)|_{[\gamma_{k+1}(r),k+1]} \leq \rho \, |\gamma'_k(r)|_{[\gamma_{k}(r),k]} + M_k, \;\; \forall r \in [0,1], \forall k \geq k_0,
\]
and by integration over $r \in [0,1]$
\[
L_{k+1}(\gamma_{k+1}) \leq \rho L_k(\gamma_k) + M_k, \; \forall k \geq k_0.
\]
By the comparison lemma \cite{Lakshmikantham02:1}, we then have that $L(\gamma_k) \leq u_k$ for $u_k$ satisfying
the linear scalar dynamics
\[
u_{k_0} = L_{k_0}(\gamma_{k_0}), \; u_{k+1} = \rho \, u_k + M_k, \, \forall k \geq k_0.
\]
Hence, $L_k(\gamma_k) \leq \rho^{k-k_0} u_{k_0} + \sum_{l=0}^{k-k_0-1} \rho^l M_{k-1-l}$. 
As in the end of the proof of Theorem \ref{thm: contractivity condition simple}, we can then choose $\gamma_{k_0}$ so 
that $L_{k_0}(\gamma_{k_0})$ is arbitrarily close to $d_{k_0}(\bar x_{0}, x_{0})$, and then use 
$d_k(\bar x_k, x_k) \leq L_k(\gamma_k)$ to conclude. 
\end{proof}

We can now make convergence assumptions on the bounding sequence $\{M_k\}_{k \geq 0}$ in \eqref{eq: bound on disturbance manifold} 
to state more concrete results.
The following corollaries follow by straightforward calculations on the sequence $u_k$ introduced at the end of the
proof of Theorem \ref{thm: contractivity under disturbance}.
\begin{corollary}	\label{cor: contraction disturbance - adj 1}
Let $1 \leq p \leq \infty$ be an integer. 
Suppose that $\{M_k\}_{k \geq 0}$ in \eqref{eq: bound on disturbance manifold} is a sequence in $\ell^p$, with norm $\|M\|_p$.
Then, with the notation and assumptions of Theorem \eqref{thm: contractivity under disturbance}, if $\rho < 1$, 
there exists a class $\mathcal K$ function $\beta: \mathbb R_+ \to \mathbb R_+$ such that
\begin{equation}	\label{eq: deviation - norm bound}
\left(\sum_{k=k_0}^\infty d_k(\bar x_k, x_k)^p \right)^{1/p}
\leq \beta(d_{k_0}(x_{0}, \bar x_{0})) + \frac{\|M\|_p}{1-\rho}, 
\end{equation}
where, for $p = \infty$, the left-hand side of the inequality is interpreted as usual as $\sup_{k \geq k_0} d_k(\bar x_k, x_k)$.
\end{corollary}
By further restricting the class of disturbances, we get slightly tighter bounds on the deviations for $p \geq 2$.
\begin{corollary}	\label{cor: contraction disturbance - adj 2}
Let $1 \leq p \leq \infty$ be an integer. 
Suppose that $\{M_k\}_{k \geq 0}$ in \eqref{eq: bound on disturbance manifold} satisfies the following
condition:
\begin{equation}	\label{eq: exponentially vanishing disturbance}
\exists K \geq 0, 1 > \alpha \geq 0, \text{and } k_0 \in \mathbb N \text{ s.t. }
M_k = \begin{cases}
0, & \text{if } k < k_0, \\
K \alpha^{k-k_0},  & \text{if } k \geq k_0.
\end{cases}
\end{equation}
Then, with the notation and assumptions of Theorem \ref{thm: contractivity under disturbance}, 
for $k \geq k_0$,
\[
d_k(\bar x_k, x_k) \leq \rho^{k-k_0} d_{k_0} (\bar x_{0}, x_{0}) + K \frac{\rho^{k-k_0} - \alpha^{k-k_0}}{\rho - \alpha}.
\]
Hence, if $\rho < 1$, 
\[
\sum_{k=k_0}^\infty d_k(\bar x_k, x_k) \leq \frac{1}{1-\rho} d_{k_0} (\bar x_{0}, x_{0}) + \frac{K}{(1-\rho)(1-\alpha)},
\]
and for any $p \geq 2$, there exists a class $\mathcal K$ function $\beta: \mathbb R_+ \to \mathbb R_+$ such that
\begin{equation}	\label{eq: deviation - refined bound}
\left( \sum_{k=k_0}^\infty d_k(\bar x_k, x_k)^p \right)^{1/p} \leq \beta(d_{k_0} (\bar x_{0}, x_{0})) + 
\frac{K}{|\rho-\alpha|} \left( \sum_{k=0}^\infty |\rho^{k} - \alpha^{k}|^p \right)^{1/p}.
\end{equation}
\end{corollary}

\begin{remark}
If the norms on $\mathsf T \mathsf X$ are given by weighted $1$ and $2$-norms as in Corollaries 
\ref{cor: contraction result 1-norm} and \ref{cor: classical contraction theorem}, then
condition \eqref{eq: contraction condition perturbed 1} corresponds to the feasibility of a family of
linear programs or LMIs, and if moreover $C$ is convex and the weight matrices in
these norms are independent of $x$, then we can replace the distances $d_k(\bar x_k, x_k)$ in 
\eqref{eq: deviation - norm bound}, \eqref{eq: deviation - refined bound}
by $|P_{[k]}(\bar x_k - x_k)|_1$ or $|P_{[k]}^{1/2}(\bar x_k - x_k)|_2$ 
as in \eqref{eq: special case constant matrices - l1}, \eqref{eq: special case constant matrices - l2}.
\end{remark}


\section{Differentially Private Observers with Output Perturbation}		\label{section: DP observers}

Let us now return to our initial differentially private observer design problem with output perturbation. 
Two adjacent measured signals $y$ and $\tilde y$ produce distinct observer state trajectories $z$ and $\tilde z$
by \eqref{eq: observer form}, such that
\begin{align}
z_{k+1} &= f_k(z_k) + h_k( z_k, y_k - g_k(z_k)),  \label{eq: observer - nominal} \\
\tilde z_{k+1} &= f_k(\tilde z_k) + h_k(\tilde z_k, y_k - g_k(\tilde z_k) + \pi_k),
\end{align}
where $\pi_k = \tilde y_k - y_k$. 
We can now attempt to choose the functions $h_k$
to design a contractive observer, while at the same time minimizing the ``gain'' of the map $\pi \to z$. 
First, contraction provides a notion of convergence for the observer. Namely, if the model 
\eqref{eq: model - 1}, \eqref{eq: model - 2} were valid under no modeling noise assumptions (zero $v, w$), 
then any the sequence $x$ satisfying \eqref{eq: model - 1}, \eqref{eq: model - 2}
would also satisfy the dynamics $\eqref{eq: observer - nominal}$ (since $y_k = g(x_k)$), and the
trajectories $x, z$ would converge exponentially toward each other, so that any initial difference between
$z_0$ and $x_0$ would eventually be forgotten.
Second, the results of Section \ref{section: disturbance effects} give us tools to bound
the sensitivity of contractive observers, i.e., the deviations between $z$ and $\tilde z$ above,
and hence a means to set the level of privacy-preserving noise using Theorem \ref{eq: basic DP mechanism}.

Given two measured signals $y$ and $\tilde y$, define the notation 
$\nu^{y, \tilde y}_k(x;r) := y_k - g_k(x) + r\pi_k = (1-r)y_k + r\tilde y_k-g_k(x)$ and
\begin{equation} \label{eq: Jacobian observer}
J^{y, \tilde y}_k(x;r) = \frac{\partial f_k}{\partial x}(x) + \frac{\partial h_k}{\partial x}(x, \nu^{y, \tilde y}_k(x;r)) 
- \frac{\partial h_k}{\partial y}(x, \nu^{y, \tilde y}_k(x;r)) \frac{\partial g_k}{\partial x}(x).
\end{equation}
The proof of the following proposition follows immediately from Theorem \ref{thm: contractivity under disturbance} 
and Remark \ref{rem: simplification of thm}.

\begin{proposition}	\label{prop: general result for observer design}
Consider the observer \eqref{eq: observer form}, and two measured signals $y, \tilde y$ producing respectively the
trajectories $z, \tilde z$, assuming the same initial condition $z_0 = \tilde z_0$ to initialize the observer.
Suppose that we have the bound
\begin{equation}
\| J^{y, \tilde y}_k(x;r) \|_{[x,k]}^{[x_+^{k,r},k+1]} \leq \rho, \quad \forall r \in [0,1], \forall x \in C, \forall k \in \mathbb N,
\end{equation}
where $J^{y, \tilde y}_k$ is defined by \eqref{eq: Jacobian observer}, $x_+^{k,r} := f_k(x) + h_k(x, \nu_k^{y,\tilde y}(x;r))$
and $C$ is a set containing $z_0$, which is forward invariant for the observer \eqref{eq: observer - nominal} for any 
input signal $(1-r)y + r \tilde y$, $r \in [0,1]$.
Suppose moreover
\begin{equation}	\label{eq: deviation bound}
\sup_{x \in C, r \in [0,1]} \left| \frac{\partial h_k}{\partial y}(x, \nu^{y, \tilde y}_k(x;r)) (\tilde y_k - y_k) \right|_{[x_+^{k,r}, k+1]} 
\leq M_k, \quad \forall k \in \mathbb N.
\end{equation}
Then, we have, for the distances $d_k$ associated to the norms $|\cdot|_{[x,k]}$
\[
d_k(z_k, \tilde z_k) \leq \sum_{l=0}^{k-1} \rho^l M_{k-1-l}.
\]
\end{proposition}

The result of Proposition \ref{prop: general result for observer design} is still quite general. To illustrate how it can be
applied and to simplify the following discussion, let us focus on the simpler case of Luenberger-type observers
\begin{equation}	\label{eq: simpler Luenberger observer}
z_{k+1} = f_k(z_k) + H_k \times (y_k - g_k(z_k)),
\end{equation}
where  
$H_k$ represents a $n \times m$ matrix to design. In other words, we set $h_k(x,y) = H_k \, y$.
Then the expression \eqref{eq: Jacobian observer} reads simply
$\frac{\partial f_k}{\partial x}(x) - H_k \frac{\partial g_k}{\partial x}(x)$
and becomes in particular independent of $r$ and $y, \tilde y$. 
Next, fix a norm $|\cdot|_\mathsf X$ on $\mathsf T \mathsf X$, independent of $x, k$,
and a $p$-norm $|\cdot|_p$ on $\mathsf Y$, and let
$\bar H_p^{\mathsf X} := \sup_{k} \|H_k\|_{\mathsf Y}^{\mathsf X}$. 
Then, in \eqref{eq: deviation bound}, we can take $M_k = \bar H_p^{\mathsf X} |y_k - \tilde y_k|_p$.
This leads to the following corollary of Proposition \ref{prop: general result for observer design},
similar to the Corollaries \ref{cor: contraction disturbance - adj 1} and
\ref{cor: contraction disturbance - adj 2}, 
which we will use next in the illustrative examples. We introduce the notation 
$\|v\|_{p, \mathsf X} := \left(\sum_{k=0}^\infty |v_k|_\mathsf X^p \right)^{1/p}$, for $1 \leq p \leq \infty$.

\begin{corollary}	\label{cor: special result for Luenberger observers}
Consider the observer \eqref{eq: simpler Luenberger observer}, and two measured signals $y, \tilde y$ producing respectively the
trajectories $z, \tilde z$, assuming the same initial condition $z_0 = \tilde z_0$ to initialize the observer.
Fix the norms $|\cdot|_{\mathsf X}$, on $\mathsf T \mathsf X$, independent of $x, k$.
Suppose that we have the bound
\begin{equation}	\label{eq: simplified bound Luenberger}
\left \| \frac{\partial f_k}{\partial x}(x) - H_k \frac{\partial g_k}{\partial x}(x) \right \|_{\mathsf X} \leq \rho, \quad \forall x \in C, k \in \mathbb N, 
\end{equation}
for some constant $\rho < 1$, 
where $C$ is a set containing $z_0$ and forward invariant for \eqref{eq: observer - nominal} for any input signal $y + (1-r) \tilde y$, $r \in [0,1]$.
Then, if the signals $y, \tilde y$ are adjacent according to \eqref{eq: adjacency relation 0}, we have, for
the same value of $p$, 
\begin{equation}	\label{eq: bound on deviation - 1}
\| z - \tilde z \|_{p,\mathsf X} \leq \frac{B_p \, \bar H_p^{\mathsf X} }{1 - \rho}.
\end{equation}
Moreover, if the signals $y, \tilde y$ are in fact adjacent according to \eqref{eq: adjacency relation}, we have
more precisely, for the same value of $p$, 
\begin{equation}	\label{eq: bound on deviation - 2}
\| z - \tilde z \|_{p, \mathsf X} \leq \frac{K \, \bar H_p^{\mathsf X}}{|\rho - \alpha|} \left( \sum_{k=0}^\infty |\rho^k - \alpha^k|^p \right)^{1/p}.
\end{equation}
\end{corollary}

\begin{remark}
For the adjacency relation \eqref{eq: adjacency relation} with $p=1$, both \eqref{eq: bound on deviation - 2}
and \eqref{eq: bound on deviation - 1} give the same upper bound $\frac{K \, \bar H_p^{\mathsf X} }{(1 - \rho)(1-\alpha)}$.
\end{remark}

In Corollary \ref{cor: special result for Luenberger observers}, the choice of $H_k$ has an impact both on $\rho$ 
and on the $\ell^p$-sensitivity bound.   
Increasing the gains $H_k$ can help decrease the contraction rate $\rho$ to obtain a more rapidly converging
observer, but at the same time it increases the sensitivity, in the sense of Section \ref{section: sensitivity}, 
and thus the level of noise necessary for differential privacy.
Hence, in general, we should try to achieve a reasonable contraction rate $\rho$ with the smallest gain possible.
We conclude this section with two more corollaries, 
describing differentially private observers with output perturbation.

\begin{corollary}		\label{eq: Laplace mechanism observer}
Let $P = \text{diag}(p)$, with $p_i > 0, 1 \leq i \leq n$,
and assume that the conditions of 
Corollary \ref{cor: special result for Luenberger observers}
are satisfied for the weighted $1$-norm $|Pv|_1 = \sum_{i=1}^n p_i |v_i|$ on $\mathsf X$.
Consider the signal 
$
\hat x_k = z_k + \xi_k,
$
where $z_k$ is computed from \eqref{eq: simpler Luenberger observer}, and 
$\xi_{k,i}$ are iid Laplace random variables with parameters $b/p_i$, for $1 \leq i \leq n$, where
\begin{equation}	\label{eq: parameter Laplace mech}
b = \frac{B_1 \sup_k \| P H_k \|_1}{\epsilon(1-\rho)}.
\end{equation}
Then this signal $\hat x_k$ is $\epsilon$-differentially private for the adjacency relation  
\eqref{eq: adjacency relation 0} with $p=1$, and for \eqref{eq: adjacency relation} 
with $p=1$ when $B_1 = \frac{K}{1-\alpha}$.
\end{corollary}

\begin{proof}
From the bound \eqref{eq: bound on deviation - 1} for $p = 1$, 
since $\|z-z\|_{1, \mathsf X} = \sum_{k=0}^\infty |P(z_k - \tilde z_k)|_1$
we deduce by Theorem \ref{eq: basic DP mechanism} that $P z_k + \zeta_k$ is 
a differentially private signal, where $\zeta_k$ has Laplace distributed iid components 
with the parameter $b$. Hence $P^{-1}(P z_k + \zeta_k)$ is also differentially private
(by resilience to post-processing\cite{LeNyDP2012_journalVersion:1}) 
and we define $\xi_k = P^{-1} \zeta_k$ in the Corollary.
\end{proof}

\begin{corollary}		\label{eq: Gaussian mechanism observer}
Let $P$ be a positive definite matrix, 
and assume that the conditions of 
Corollary \ref{cor: special result for Luenberger observers}
are satisfied for the weighted $2$-norm $| P^{1/2} v |_2$ on $\mathsf X$.
Consider the signal
$
\hat x_k = z_k + \xi_k,
$
where $z_k$ is computed from \eqref{eq: simpler Luenberger observer}, 
and $\xi_k$ is a Gaussian white noise with covariance matrix $\sigma^2 P^{-1}$, where
$
\sigma = \kappa_{\delta,\epsilon} K_2 \sup_k \| P^{1/2} H_k \|_2.
$
Then this signal $\hat x_k$ is $(\epsilon,\delta)$-differentially private for the adjacency relation 
\eqref{eq: adjacency relation 0} with $p=2$ if $K_2 = B_2/(1-\rho)$, 
and for the adjacency relation \eqref{eq: adjacency relation} with $p=2$ if
$K_2 = \frac{K}{|\rho - \alpha|} \left(\sum_{k \geq 0} (\rho^k - \alpha^k)^2 \right)^{1/2}$.
\end{corollary}
\begin{proof}
From the bounds \eqref{eq: bound on deviation - 1} or \eqref{eq: bound on deviation - 2}, we deduce 
by Theorem \ref{eq: basic DP mechanism} that $P^{1/2} z_k + \zeta_k$ is a differentially private signal, 
where $\zeta_k$ is a Gaussian white 
noise with covariance matrix $\sigma^2 I$. 
Hence $P^{-1/2}(P^{1/2} z_k + \zeta_k)$ is also 
differentially private (by resilience to post-processing\cite{LeNyDP2012_journalVersion:1}) and we 
define $\xi_k = P^{-1/2} \zeta_k$ in the Corollary.
\end{proof}

Corollaries \ref{eq: Laplace mechanism observer} and \ref{eq: Gaussian mechanism observer} give 
two differentially private mechanisms with output perturbation, provided we can design the 
matrices $\bar H_k$ to verify the assumptions of 
Corollary \ref{cor: special result for Luenberger observers}
with the (weighted) $1$- or $2$-norm on $\mathsf X$. 
The next section discusses application examples for the privacy-preserving observer design methodology.


\section{Examples}
\subsection{Estimating Link Formation Preferences in Dynamic Social Networks}	\label{section: DSBM}

Statistical studies of networks have intensified tremendously in recent years, with one motivating
application being the emergence of online social networking communities. 
In this section we focus on a recently proposed state-space model\cite{XuHero14_dynamicNetworks:1}
describing the dynamics of link formation in networks, called the Dynamic Stochastic Blockmodel. 
It combines a linear state-space model for the underlying dynamics of the network and the
classical stochastic blockmodel of Holland et al. \cite{Holland83_blockmodel:1}, resulting in a nonlinear measurement 
equation. 
Examples of applications of this model include mining email and cell phone databases 
\cite{XuHero14_dynamicNetworks:1}, which obviously contain privacy-sensitive data.

Consider a set of $n$ nodes. Each node corresponds to an individual and can belong to one of $N$ classes. 
Let $\theta^{ab}_{k}$ be the probability of forming an edge at time $k$ between a node in class $a$ and a 
node in class $b$, and let $\theta_k$ denote the vector of probabilities $[\theta^{ab}_k]_{1 \leq a,b \leq N}$.
For example, edges could represent email exchanges or phone conversations. 
Edges are assumed to be formed independently of each other according to $\theta_k$. 
Let $y^{ab}_k=\frac{m^{ab}_k}{n^{ab}}$ be the observed density of edges between classes $a$ and $b$,
where $m^{ab}_k$ is the number of observed edges between classes $a$ and $b$ at time $k$, and 
$n^{ab}$ is the maximum possible number of edges between these two classes. 
For simplicity, we assume that the quantities $n^{ab}$ are publicly known (this is the case, for example, if the class of each node 
is public information), and we focus on the problem of estimating the parameters $\theta^{ab}_k$ by using the signals
$y_k^{ab}$. This corresponds to the ``a priori'' blockmodeling setting\cite{Holland83_blockmodel:1, XuHero14_dynamicNetworks:1}.
The links formed between specific nodes constitute private information however, so directly releasing 
$m^{ab}_k$ or $y^{ab}_k$ or an estimate of $\theta_k$ based on these quantities is not allowed. 

If $n^{ab}$ is large enough, previous work has argued\cite{XuHero14_dynamicNetworks:1} using 
the Central Limit Theorem that an approximate model where $y^{ab}_k$ is Gaussian is justified, so that
\begin{equation}	\label{eq: obs model network 1}
y_k = \theta_k + v_k,
\end{equation}
where $v_k$ is a Gaussian noise vector with diagonal covariance matrix $V_k$ (whose entries theoretically
should depend on $\theta_k$, but this aspect is neglected in the model). 
Rather than defining a dynamic model for $\theta_k$, whose entries are constrained to be between $0$ and $1$,
let us redefine the state vector to be the so-called logit of $\theta_k$, denoted $\psi_k$, with entries 
$\psi^{ab}_k = \ln \frac{\theta^{ab}_k}{1-\theta^{ab}_k}$, which are well defined for $0 < \theta_k^{ab} < 1$. 
The dynamics of $\psi_k$ is assumed to be linear  
\begin{align}	\label{eq: dynamic model network}
\psi_{k+1} = F \psi_k + w_k,
\end{align}
for some known matrix $F$, and for noise vectors $w_k$ assumed to be iid Gaussian 
with known covariance matrix $W$ \cite{XuHero14_dynamicNetworks:1}.
The observation model (\ref{eq: obs model network 1}) now becomes
\begin{equation}	\label{eq: observation model network 2}
y_k = g(\psi_k) + v_k,
\end{equation}
where the components of $g$ are given by the logistic function applied to each entry of $\psi$, i.e., 
\[
g^{ab}(\psi_k) = \frac{1}{(1+e^{-\psi^{ab}_k})}.
\] 

An Extended Kalman Filter (EKF) is proposed in \cite{XuHero14_dynamicNetworks:1}
to estimate $\psi$, but we pursue here a deterministic observer design to illustrate 
the ideas discussed in the previous sections.
Hence, for simplicity we consider an observer  of the form
\[
\hat \psi_{k+1} = F \hat \psi_k + H (y_k - g(\hat \psi_k)) = (F \hat \psi_k - H g(\hat \psi_k)) + H y_k,
\]
with $H$ a constant square gain matrix.
To enforce contraction as in Corolloary \ref{cor: special result for Luenberger observers},
we should choose $H$ so that 
$
\|F - H G(\psi)\| \leq \rho,
$
where $G(\psi)$ is the Jacobian of $g$ at $\psi$. Note that $G(\psi)$ is a square and diagonal matrix with entries
$
G_{ii}(\psi) = \frac{e^{-\psi^i}}{(1+e^{-\psi^i})^2},
$
with $i$ indexing the pairs $(a,b)$.
The only nonlinearity in the model (\ref{eq: dynamic model network}), (\ref{eq: observation model network 2}) 
comes from the observation model (\ref{eq: observation model network 2}). 

To further simplify the following discussion, let us assume that $F$ is also diagonal (an assumption also made
in previous work\cite{XuHero14_dynamicNetworks:1}, where the coupling between components occurs only 
through the non-diagonal covariance matrix $W$).
In this case, the systems completely decouple into scalar systems, and it is natural to choose $H$ to be 
diagonal as well. The observer for one of these scalar system takes the form
\begin{align}	\label{eq: scalar version observer}
z_{k+1} = f z_k + h \times \left( y_k - \frac{1}{1+e^{-z_k}} \right) = f z_k - \frac{h}{1+e^{-z_k}} + h y_k,
\end{align}
where $h \in \mathbb R$ is the observer gain to set, $f \in \mathbb R_+$,
$z_k \in \mathbb R$ is one component $(a,b)$ of $\hat \psi_k$ and $y_k$ now represents just the corresponding 
scalar component of the measurement vector as well. Since the state space $\mathsf X$ is now $\mathbb R$,
the norm $| \cdot |_\mathsf X$ is simply the absolute value.
The contraction condition \eqref{eq: simplified bound Luenberger} reads, for some $0 < \rho < 1$,
\begin{align}
-\rho &\leq f - \frac{h \, e^{-z}}{(1+e^{-z})^2} \leq \rho \\
\text{i.e., } \; f-\rho &\leq \frac{h \, e^{-z}}{(1+e^{-z})^2} \leq f+\rho. \label{eq: condition on gain} 
\end{align}
Now note that $0 \leq \frac{e^{-z}}{(1+e^{-z})^2} \leq \frac{1}{4}$ for all $z$. 
Hence, by taking $h \leq 4 (f+\rho)$, the right inequality (\ref{eq: condition on gain}) is satisfied.
To satisfy the left inequality, if $f < 1$, we could potentially take $\rho \geq f$, although the estimation performance 
might not necessarily be satisfying in this case.
Alternatively, if $f \geq 1$ or if we want to achieve a smaller contraction parameter $\rho$ than the value of $f$, we can enforce 
the left inequality on a subset of the state-space. Namely, for $-a \leq z \leq a$, we have 
$\frac{e^{-z}}{(1+e^{-z})^2} \geq \frac{e^{-a}}{(1+e^{-a})^2}$. In this case, for $\rho < f$, 
by taking $h \geq (f-\rho)e^a(1+e^{-a})^2$, the left hand side of (\ref{eq: condition on gain}) is also satisfied.

Suppose for example that $f = 1$ in the dynamics \eqref{eq: scalar version observer}, so that \eqref{eq: dynamic model network}
describes a Gaussian random walk, and that the adjacency relation considered is \eqref{eq: adjacency relation}.
By Corollary \ref{eq: Laplace mechanism observer}, we can publish an $\epsilon$-differentially private estimate of $\psi$ 
by computing $z_k$ using \eqref{eq: scalar version observer} and adding Laplace noise to it with parameter 
$b = K h / (\epsilon (1-\rho) (1-\alpha))$. Small noise requires small values of $h$ and of $\rho$. 
Since we must take $\rho < 1$, we cannot enforce the left inequality
of \eqref{eq: condition on gain} for all values of $z$. Suppose then that we want to design a privacy-preserving 
observer assuming that $\theta$ remains in the interval $[0.1,0.9]$, or equivalently $\psi \in [-2.197,2.197]$ approximately. 
In this interval, we have
$
0.09 \leq \frac{e^{-\psi}}{(1+e^{-\psi})^2} \leq \frac{1}{4},
$
and so $\rho$ and $h$ must also satisfy
\begin{equation}	\label{eq: numerical inequalities DSBM}
\frac{f-\rho}{0.09} \leq h \leq 4(f+\rho), \quad \text{i.e., } \frac{1-\rho}{0.09} \leq h \leq 4(1+\rho).
\end{equation}
Note in particular that the factor $h/(1-\rho)$ also appearing in the parameter $b$ is lower bounded by $1/0.09 \approx 11.1$.  
We should then set $h = (1-\rho)/0.09$, satisfying the left inequality in \eqref{eq: numerical inequalities DSBM} with equality, 
for the value of the contraction parameter $\rho$ that we want to achieve.
For example, for faster observer convergence we should try to achieve the lowest possible value of $\rho$, although this might
amplify the steady-state variance due to measurement noise. 
The inequalities \eqref{eq: numerical inequalities DSBM}
can only be satisfied for $\rho \gtrapprox 0.47$, a contraction parameter that can then be achieved by taking $h \approx 5.88$.

\begin{figure}
\centering
\includegraphics[width=0.8\linewidth]{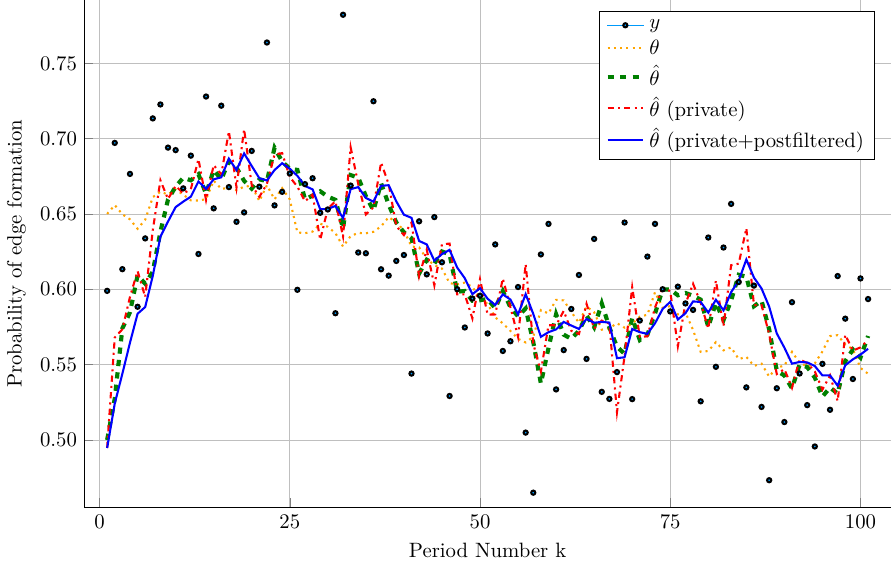}
\caption{Sample path of the estimate of the edge formation probability $\theta_k^{ab}$, for some classes $(a,b)$. 
The measured edge density is generated from
one component of the model (\ref{eq: obs model network 1}), (\ref{eq: dynamic model network}) with
$f=1$ and $w_k, v_k$ iid Gaussian random variables with zero mean and standard deviation $0.03$ and
$0.04$ respectively. 
The trajectory $\theta$ (dotted line) starts at the value $0.65$, and the observers are all initialized at the value $0.50$.
The upper bound $\rho$ on the contraction rate of the observer \eqref{eq: scalar version observer}
is set to $\rho = 0.9$, providing a good tradeoff between convergence speed and steady-state variance (see
green dashed curve for the non-private observer), with corresponding gain $h = 1.11$.
The dot-dashed line shows $1/(1+\exp(-\tilde z_k))$ as our private estimate of $\theta_k$, where $\tilde z_k$ is 
a $\ln(3)$-differentially private estimate of $\psi_k$ (hence, the estimate of $\theta_k$ is also $\ln(3)$-differentially private), 
obtained by the Laplace mechanism, for the adjacency relation (\ref{eq: adjacency relation}) with parameter 
values $K = 3 \times 10^{-3}$, $\alpha = 0.25$.
We also show a differentially private estimate obtained after further post-filtering, as explained in the main text.}  
\label{fig: blockmodel estimation}
\end{figure}

Figure \ref{fig: blockmodel estimation} illustrates the behavior of the privacy-preserving observer,
when the privacy parameters are $\epsilon = \ln(3), \delta = 0$ and
$K = 3 \times 10^{-3}$ and  $\alpha = 0.25$ in \eqref{eq: adjacency relation}.
That is, for the pair of classes $(a,b)$ under consideration, we want
to provide a differential privacy guarantee making it hard to detect a transient variation in the number of edges,  
as long as this variation represents initially at most $0.3\%$ of all the edges between classes $a$ and $b$,
and subsequently decreases at least geometrically with rate $1/4$. Concretely, if edges represent phone conversations
for example, this means that if an individual in class $a$ suddenly increases his call volume with class $b$ but 
by an amount representing less than a proportion $K$ of all calls between $a$ and $b$, and subsequently reduces this temporary 
activity at rate $\alpha$, then an adversary having access  
to a differentially private estimate of $\theta_k^{ab}$ .
can only achieve a low probability of correctly detecting this event\cite{Wasserman:DPstat}.

As explained in Figure \ref{fig: input and output perturbation}, it can be useful to further filter the differentially private
signal produced above, since this signal exposes directly the privacy-preserving noise. In this case, one can interpret 
the private estimate $\tilde z_k = z_k + \xi_k$, with $\xi$ the Laplace noise as in Corollary \ref{eq: Laplace mechanism observer}, 
as a noisy measurement of $\psi$, now with a trivial, linear measurement model in contrast to \eqref{eq: observation model network 2}.
A possible simple post-filter smoothing $\tilde z_k$ can then be the linear observer 
\[
\hat \psi_{k+1} = f \hat \psi_k + k_{\text{post}} (\tilde z_k - f \hat \psi_k),
\]
and Figure \ref{fig: blockmodel estimation} also represents $\hat \theta_k = g(\hat \psi_k)$ for the gain value $k_{\text{post}} = 0.4$.

\subsection{Syndromic Surveillance}	\label{ref: syndromic surveillance}

Syndromic surveillance systems monitor health related data in real-time in a population 
to facilitate early detection of epidemic outbreaks \cite{Lawson05_syndromicSurveillance:1}. 
In particular, recent studies have shown the correlation between certain non-medical data,
e.g., search engine queries related to a specific disease, and the proportion of individuals 
infected by this disease in the population \cite{Ginsberg09_syndromic:1}. 
Although time series analysis can be used to detect abnormal patterns in the collected 
data \cite{Lawson05_syndromicSurveillance:1}, 
here we focus on a model-based filtering approach \cite{Skvortsov12_syndromicSurveillance:1},
and develop a differentially private observer for a $2$-dimensional epidemiological model.

The following SIR model of Kermack and McKendrick 
\cite{Kermack1927_epidemics:1, Brauer08_epidemiology:1} models the evolution of an epidemic in a population 
by dividing individuals into 3 categories: susceptible (S), i.e., individuals who might become infected if exposed; 
infectious (I), i.e., currently infected individuals who can transmit the infection; and recovered (R) individuals, who are
 immune to the infection.
A simple version of the model in continuous-time includes bilinear terms and reads
\begin{align*}
\frac{ds}{dt} &= -\mu \mathcal R_o i s \\
\frac{di}{dt} &= \mu \mathcal R_o i s - \mu i.
\end{align*}
Here $i$ and $s$ represent the proportion of the total population in the classes $I$ and $S$. 
The last class $R$ need not be included in this model because we have the constraint $i+s+r=1$. 
The parameter $\mathcal R_o$ is called the basic reproduction number and represents 
the average number of individuals infected by a sick person. The epidemic can
propagate when $\mathcal R_o > 1$. The parameter $\mu$ represents the rate at which infectious
people recover and move to the class $R$. More details about this model can be found in \cite{Brauer08_epidemiology:1}.

Discretizing this model with sampling period $\tau$, we get the discrete-time model
\begin{align}
s_{k+1} &= s_k - \tau \mu \mathcal R_o i_k s_k + w_{1,k} = f_1(s_k,i_k) + w_{1,k}  \label{eq: SIR model eq 1} \\
i_{k+1} &= i_k + \tau \mu i_k (\mathcal R_o s_k - 1) + w_{2,k} = f_2(s_k,i_k) + w_{2,k},  \label{eq: SIR model eq 2}
\end{align}
where we have also introduced noise signals $w_1$ and $w_2$ in the dynamics.
We assume here for simplicity that we can collect syndromic data providing a noisy measurement
of the proportion of infected individuals, .i.e.,
\[
y_k = i_k + v_k,
\]
where $v_k$ is a noise signal. We can then consider the design of an observer of the form
\begin{align*}
\hat s_{k+1} &= f_1(\hat s_k, \hat i_k) + h_1 (y_k - \hat i_k) \\
\hat i_{k+1} &= f_2(\hat s_k, \hat i_k) + h_2 (y_k - \hat i_k).
\end{align*}
We define the Jacobian matrix of the system (\ref{eq: SIR model eq 1}), (\ref{eq: SIR model eq 2})
\[
F(s,i) = I_2 + \tau \mu \mathcal R_o \begin{bmatrix} -i & -s \\ i & s - 1/\mathcal R_o \end{bmatrix},
\]
as well as the gain matrix $H = [h_1, h_2]^T$ and observation matrix $C = [0,1]$.
Here, we design a differentially private observer with Gaussian noise using Corollolary \ref{eq: Gaussian mechanism observer},
for the adjacency relation \eqref{eq: adjacency relation} with $p=2$.

Following Corollary \ref{cor: classical contraction theorem}, 
the contraction rate constraint \eqref{eq: simplified bound Luenberger}  
for a 2-norm on $\mathbb R^2$ weighted by a matrix $P \succ 0$ 
is equivalent to the family of inequalities, for all $(s,i)$ in the region of $[0,1]^2$ where we want to show contraction
\begin{align*}
&(F(s,i) - HC)^T P (F(s,i) - HC) \preceq \rho^2 P \\
& F_x^T P F_x - F_x^T P H C - C^T H^T P F_x + C^T H^T P H C \preceq \rho^2 P,
\end{align*}
where we used $F_x := F(s,i)$ to simplify the notation. Defining the new variable $X = PH$, this can
be rewritten
\[
F_x^T P F_x - F_x^T X C - C^T X^T F_x + C^T X^T P^{-1} X C \preceq \rho^2 P,
\]
which, using the Schur complement, is equivalent to the family of LMIs
\begin{equation}	\label{eq: LMI contraction rate}
\begin{bmatrix}
\rho^2 P - F_x^T P F_x + F_x^T X C + C^T X^T F_x  & C^T X^T \\ X C & P 
\end{bmatrix} \succeq 0,
\end{equation}
for all $x=(s,i)$ in the region where we want to prove contraction. If we can find $P, X$ satisfying
these inequalities, we recover the observer gain vector simply as $H = P^{-1} X$.

For a given value of $\rho$, the covariance matrix of the Gaussian noise in Corollary \ref{eq: Gaussian mechanism observer} 
is proportional to $\|P^{1/2} H\|_2^2 P^{-1} = (H^T P H) P^{-1} = (X^T P^{-1} X) P^{-1}$, and hence it is advantageous to minimize
a function of this matrix. Note that $X^T P^{-1} X$ is a scalar. Minimizing $(X^T P^{-1} X) \, \text{Tr}(P^{-1})$ does not appear to
directly lead to an efficiently solvable optimization problem, but as a proxy we can choose to minimize instead the sum $X^T P^{-1} X + \nu \text{Tr}(P^{-1})$,
for some tuning parameter $\nu$.
After taking Schur complements, this leads to the following semidefinite program, 
for a given value of the contraction parameter $\rho$ 
\begin{align*}
\min_{\Sigma \succeq 0, \lambda \geq 0, P \succeq 0, X} \quad & 
\lambda + \nu \, \text{Tr}(\Sigma) \\
\text{subject to } & 
\begin{bmatrix} \lambda & X^T \\ X & P \end{bmatrix} \succeq 0, 
\;\; 
\begin{bmatrix} \Sigma & I_2 \\ I_2 & P \end{bmatrix} \succeq 0, \text{ and } \eqref{eq: LMI contraction rate}.
\end{align*}
Alternatively, one can minimize $\lambda \text{Tr}(\Sigma)$ for fixed values of $\lambda$ subject to the constraints above and perform 
a one-dimensional search for a minimizing value of $\lambda$.

\begin{figure}
\centering
\includegraphics[width=0.6\linewidth]{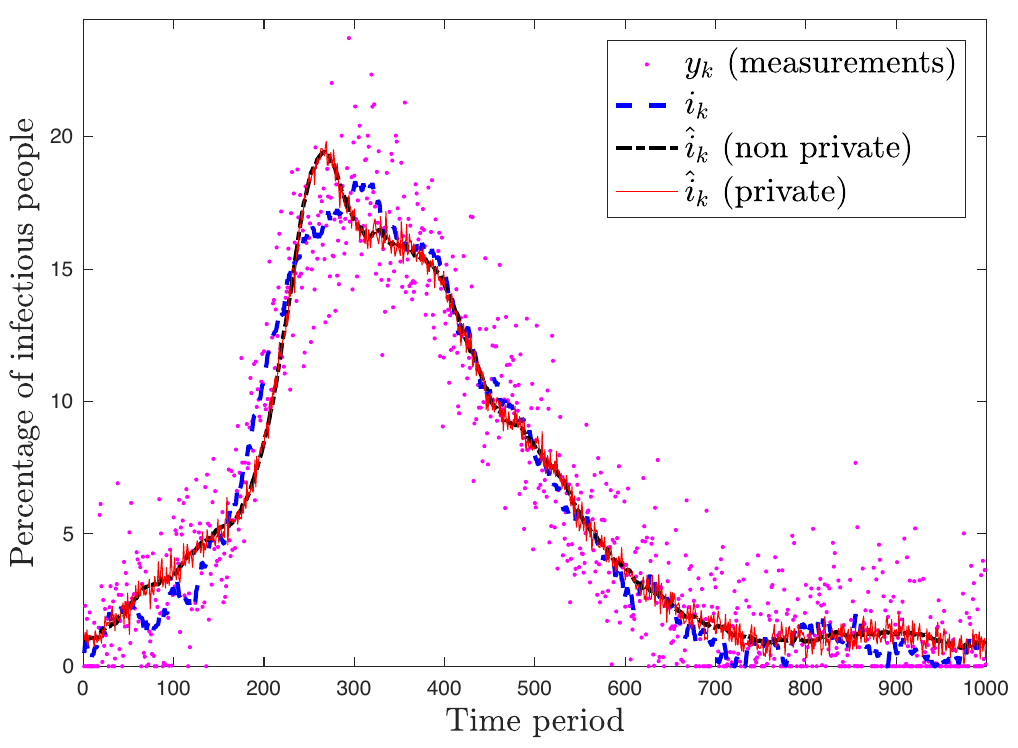}
\caption{Sample path of the estimate of the percentage of infectious people over time produced by the observer. 
The standard deviations for the dynamics and measurement noise were set to $\sigma_{w_k} I_2 = 0.005 \sqrt{\tau} I_2$ and $\sigma_{v_k} = 0.02$ 
respectively. The signals were truncated to maintain positive values for $i, s, y$ in the simulation. 
The true proportion of infectious people starts at $0.5\%$, whereas the estimate used to initialize
the observer is $1\%$.
The output of the differentially private observer is not filtered.
}
\label{fig: epidemic}
\end{figure}

\begin{eexample}
Let us assume $\mu=0.1$, $\mathcal R_o = 2$, $\tau = 0.1$, $K = 10^{-3}, \alpha=0.25$ in (\ref{eq: adjacency relation}),
and $\epsilon = 2$, $\delta = 0.05$. That is, we wish to provide a $(2,0.05)$-differential privacy guarantee for 
maximum deviations of $0.1\%$ (see the discussion in the previous subsection).  
Although not a perfectly rigorous contraction certificate, we sample the continuous set of constraints (\ref{eq: LMI contraction rate})
by sampling the set $\{(s,i) | 0.01 \leq i \leq 0.25, 0.01 \leq s \leq 1-i \}$ at the values of $s,i$ multiple of $0.01$, to
obtain a finite number of LMIs. 
A more rigorous approach to enforce these 
constraints could make use of sum-of-squares programming \cite{Aylward08_SOScontraction:1}.
Following the procedure above, for the choice $\rho = 0.996$, we obtain 
the observer gain 
$H = [3.9304;0.2003]$
and the covariance matrix $\Sigma$ with $\Sigma^{1/2} = \begin{bmatrix} 691 & 22 \\ 22 & 17 \end{bmatrix} \times 10^{-4}$
for the privacy-preserving Gaussian noise.
Sample trajectories of the non-private and private (non-smoothed) estimates of $i$ are shown on Fig. \ref{fig: epidemic}.
\end{eexample}

\section{Conclusion}
\label{section: conclusion}

This paper introduces a design methodology for nonlinear observer design, 
which provides differential privacy guarantees when the measured signals are privacy sensitive,
by perturbing the published output signal of the observer.
Tools from contraction analysis are used both to enforce convergence of the observer and
to set the level of output noise necessary in order to provide the differential privacy guarantee. 
More concretely, we bound the sensitivity of the observers by leveraging a robustness property of contractive systems.
The observer design methodology is illustrated through two examples where we construct estimators 
for models with nonlinear dynamics or measurements. 

\bibliographystyle{ieeetr}
\bibliography{./biblio}

\end{document}